\documentstyle[twocolumn,prl,aps, graphicx, epsf, psfig]{revtex}
\input{epsf}
\begin{document}
\title
{ Polarization transfer in
the  $^{4}$He$(\vec e,e^{\prime} \vec p)
^{3}$H reaction}
\author{
S.~Dieterich,$^{1}$
P.~Bartsch,$^{2}$
D.~Baumann,$^{2}$
J.~Bermuth,$^{3}$
K.~Bohinc,$^{2,4}$
R.~B\"ohm,$^{2}$
D.~Bosnar,$^{2,a}$
S.~Derber,$^{2}$
M.~Ding,$^{2}$
M.~Distler,$^{2}$
I.~Ewald,$^{2}$
J.~Friedrich,$^{2}$
J.M.~Friedrich,$^{2,b}$
R.~Gilman,$^{1,5}$ 
C.~Glashausser,$^{1}$ 
M.~Hauger,$^{6}$
P.~Jennewein,$^{2}$
J.~Jourdan,$^{6}$
J.J.~Kelly,$^{7}$
M.~Kohl,$^{8}$
A.~Kozlov$^{2}$,
K.W.~Krygier,$^{2}$
G.~Kumbartzki,$^{1}$
J.~Lac,$^{1,c}$
A.~Liesenfeld,$^{2}$
H.~Merkel,$^{2}$
U.~M\"uller,$^{2}$
R.~Neuhausen,$^{2}$
Th.~Pospischil,$^{2}$
R.~D.~Ransome,$^{1}$
D.~Rohe,$^{6}$
G.~Rosner,$^{2,d}$
H.~Schmieden,$^{2}$
M.~Seimetz,$^{2}$
I.~Sick,$^{6}$
S.~Strauch,$^{1}$
J.M.~Udias,$^{9}$
J.R.~Vignote,$^{9}$
A.~Wagner,$^{2}$
Th.~Walcher,$^{2}$
G.~Warren,$^{6}$
M.~Weis$^{2}$
}
\address{
\vspace{0.1cm}
$^{1}$Rutgers, The State University of New Jersey, Piscataway, New Jersey, USA\\
$^{2}$Institut f\"ur Kernphysik, Universit\"at Mainz, Mainz, Germany\\
$^{3}$Institut f\"ur Physik, Universit\"at Mainz, Mainz, Germany\\
$^{4}$Institut Jozef Stefan, Ljubljana, Slovenia\\
$^{5}$Thomas Jefferson National Accelerator Facility, 
Newport News, Virginia, USA\\
$^{6}$Universit\"at Basel, Basel, Switzerland\\
$^{7}$University of Maryland, College Park, Maryland, USA\\
$^{8}$Institut f\"ur Kernphysik, Technische Universit\"at Darmstadt,
Darmstadt, Germany\\
$^{9}$Universidad Complutense de Madrid, Madrid, Spain\\
}
\vspace{0.1cm}
\date{\today}
\maketitle
\begin{abstract}
Polarization transfer in the $^4$He$(\vec e,e^\prime\vec p)^3$H
reaction at a $Q^2$ of 0.4 (GeV/c)$^2$ was measured at the Mainz
Microtron MAMI.  The ratio of the transverse to the longitudinal
polarization components of the ejected protons was
compared with the same ratio for elastic $ep$ scattering.  The results
are consistent with a recent fully relativistic calculation
which includes a predicted medium modification of the proton form
factor based on a quark-meson coupling model.
\end{abstract}

\vspace{0.2cm}
\vspace{0.35cm}

A long standing question in nuclear physics is the effect of the
nuclear medium on the properties of the nucleon.  The close proximity
of nucleons in the nucleus would  lead one to expect effects
on the spatial distribution of the 
nucleon's constituent particles.  However, experimentally
distinguishing changes of the nucleon structure from other 
conventional 
nucleus-related effects, such as meson-exchange currents (MEC), isobar
configurations (IC), and final state interactions (FSI) has proven
difficult.  The form factor of a bound nucleon is not directly
observable; it must be inferred from calculations which 
predict how a modification of the  form factor
will affect  measurable quantities such as cross sections or
polarizations.

In this paper we report on the first measurement of polarization
transfer in the $^4$He$(\vec{e},e'\vec{p})^{3}$H reaction in
quasielastic parallel kinematics.  Polarization provides a sensitive test
of any model and should be more sensitive to changes
in the form factor than cross section measurements. 

Several recent calculations \cite{Udias,Kelly99,Udias00} have demonstrated the
importance of dynamic enhancement of lower components of Dirac
spinors (spinor distortions) 
for the $(e,e'p)$ reaction.  The $R_{LT}$ response function is
 sensitive to distortion of the bound-state spinor while
recoil polarization is more sensitive to distortion of the ejectile
spinor.  The relativistic calculations of Udias {\it et al.} provide 
excellent descriptions of $A_{TL}$ in $^{16}$O$(e,e'p)$ 
 \cite{Gao00} 
and the induced 
polarization for $^{12}$C$(e,e'\vec p)$ \cite{Woo}.  
The sensitivity of recoil polarization to 
possible density
dependence of nucleon form factors was investigated first by Kelly
\cite{Kelly97} using
a local density approximation for the current operator and an
effective momentum approximation (EMA) for spinor distortion.  
Udias  then
performed a fully relativistic calculation which shows that the accuracy
of the EMA for the recoil polarization is better than 1\% for missing
momentum $p_{m} < 100$ MeV/c.  
Both groups have shown that recoil polarization for modest
$p_m$ is relatively insensitive to gauge and Gordon ambiguities and to
variations of the optical potential and have concluded that polarization
transfer provides a promising probe of density-dependent modifications
of nucleon electromagnetic form factors.

Polarization transfer was first
used to study  nuclear medium effects in
deuterium \cite{Eyl95,Mil98,Bark99}.  Within
statistical uncertainties, no evidence of medium
modifications was found. Malov {\it et al.} \cite{Mal00} made the first
measurement of polarization transfer in a complex nucleus, $^{16}$O.
Their results were consistent with predictions of relativistic
calculations,  with limited  statistical precision.

Cross section data indicate only upper limits on possible
 modifications of the form factors in the nucleus.
The  limits  come
primarily from quasielastic electron
scattering with separation of the longitudinal  and transverse
response functions \cite{Zgh94,Schia89,Ducret93,Bl95} 
and from $y$-scaling \cite{Sick,Day90,Day87} of inclusive electron
scattering.  
In the Q$^{2}$ range of 0.1 to 0.5 (GeV/c)$^{2}$ $L/T$ separations
limit modifications to 3\% for the magnetic and 5-10\% for the
electric form factor.
The limits from $y$-scaling at higher $Q^{2}$,
in the range of 1 to 5  (GeV/c)$^{2}$, are about the same.  

While the data  exclude substantial form factor
modification, especially of the magnetic form factor, recent
theoretical work predicts modifications within the experimental limits
\cite{Adelaide,Frank,Meissner,Cheon}.  Lu {\it et al.}  \cite{Adelaide},
using a quark-meson coupling model (QMC), and Frank {\it et al.}
\cite{Frank}, using a light front constituent quark model, both predicted
changes of a few percent in the form factors.  Lu calculated the
change 
for both $^{16}$O and $^{4}$He and found little difference in the size
of the modification.  
We shall later examine the effect on the predicted polarization
of  the QMC modification.

In the case of electron-nucleon
scattering, there is a direct relationship between the form factors
and the polarization transfer components \cite{Arnold}:

\begin{equation}
   \frac{G_E}{G_M}
  = -\frac{P_x^{\prime}}{P_z^{\prime}} \cdot \frac{E+E^\prime}{2m_N} \tan(\theta/2)
\label{eq:gegm}
\end{equation}

\noindent where $E$ and $E^\prime$ are the energies of the incident
and scattered electron,
$\theta$ is the electron scattering angle,  $m_N$ is the nucleon
mass, and  the longitudinal and
transverse polarization transfer observables are
 $P_z^\prime$ and $P_x^\prime$, respectively. 
The relation in Eq.~(\ref{eq:gegm}) was recently used to extract
$G_{E}/G_{M}$ for the proton \cite{Mil98,Jones}.  This relationship is only
approximately correct for electron scattering from a bound nucleon; one must
calculate the expected polarization ratio in the context of some
model. 

We use the
coordinate system with unit vectors pointing in the direction of
the three momentum transfer $\hat{z} = \hat{q}$, normal to the
electron scattering plane $\hat{y} = (\hat{k}_i \times \hat{k}_f)/
|\hat{k}_i \times \hat{k}_f|$, and transverse $\hat{x} = \hat{y}
\times \hat{z}$, where the initial and final electron 
 momenta are $\vec k_i$ and $\vec k_f$.  
The results are presented
 in terms of  ${P_x^\prime}$ and ${P_z^\prime}$,
the projections of the transferred polarizations on these axes.

The advantage of using polarization is that the polarizations do not depend on
the target thickness or total current;
 the beam polarization and analyzing power both cancel in the
ratio of polarizations.  
 The difference between the asymmetries measured with
positive and negative beam helicity cancels instrumental
asymmetries to first order.  The only significant experimental
systematic uncertainty is the determination of the spin precession in
the spectrometer. 

This experiment was done at MAMI at the Johannes
Gutenberg-Universit\"at, Mainz, Germany, using
the spectrometer facilities
 of the A1 collaboration Ref.~\cite{Blom98}.  A proton focal
plane polarimeter (FPP) was installed in Spect.~A \cite{Posp00}.
  The beam energy
was 854.5 MeV.  The nominal settings  were 625 MeV/c
 central momentum at $50.24^{\circ}$ for the electron arm
(Spect.~B) and 660 MeV/c and $46.56^{\circ}$
for the proton arm (Spect.~A).  The data covered the $Q^{2}$ range
of 0.35 to 0.42 (GeV/c)$^2$.
To measure the polarization 
ratio for the free proton and to study systematic effects,  $\vec{e}p$ elastic
scattering data were taken at the same nominal settings as for
the $^{4}$He measurement  except that  the central angle for the
proton arm was set to
$48.16^{\circ}$ in order to better
match the kinematic acceptance.  

The FPP includes  a graphite analyzer with a thickness of 11.9
g/cm$^{2}$ (7 cm).  The spectrometer vertical drift
chambers  serve as the tracking detectors before the analyzer and
two horizontal drift chambers track scattered protons after the
analyzer.  The basic design of the FPP is similar to several used
previously \cite{Polar}. For further details, see Ref.~\cite{Posp00}.

The helium target consisted of a gas cell 8 cm long, at a temperature
 of 19 K  and pressure of 19 bar. The target thickness was about 250
mg/cm$^{2}$.  The same cell was used for liquid hydrogen,
with a thickness of about 560
mg/cm$^{2}$, in order
to minimize any systematic differences.  
The beam current used for hydrogen was typically 0.5 $\mu$A,
set  by the data acquisition rate,
and for helium 14 $\mu$A.  
 The beam polarization was approximately 75\%, as determined
from the recoil polarization measured for hydrogen using the analyzing power
from Ref.~\cite{Anal}.
The invariant mass resolution was approximately 0.8 MeV,
which allowed clear separation of the $^{3}$H$p$ 
final state from the $^{2}$H$pn$
and $ppnn$ final states.

In the data analysis the criteria for selected events included,
among others, tests on spectrometer acceptance, target geometry,
missing mass, and FPP polar scattering angle ($7^\circ < \theta_{c} <
35^\circ$). The physical quantities of interest, $P_x^\prime$ and
$P_z^\prime$, were determined by means of the maximum likelihood
technique, utilizing the azimuthal distribution of the protons
scattered from the graphite analyzer

\begin{equation}
I=I_{0}\left[1+\epsilon_{y}\cos (\phi_{c})+\epsilon_{x}\sin(\phi_{c})\right].
\label{eq:fppasym}
\end{equation}

\noindent The asymmetries $\epsilon_x$ and $\epsilon_y$ are proportional
to the analyzing power and to the proton's
polarizations perpendicular to its momentum as it enters the analyzer
and  are linear functions of the proton's 
polarization components at the target. The  relationship is given
by a rotation which takes into account the change of coordinate system
and the proton spin precession \cite{Posp00} 
in the spectrometer's magnetic fields
and is calculated on an event by event basis.

The systematic uncertainty in the determination of
$ {P_x^{\prime}}/{P_z^{\prime}}$ 
can be estimated by introducing artificial shifts in various
parameters and finding the effect on the ratio.  The total
systematic uncertainty was estimated to be  $\pm 0.03$ on the
ratio, for both helium and hydrogen.

\vbox{
\begin{table}
\caption{Polarization ratios with statistical and estimated
systematic uncertainties. The world average, last row, is derived from
measurements of $G_E$ and $G_M$.}
\begin{tabular}{@{\hspace{10mm}}lll@{\hspace{10mm}}}
\multicolumn{1}{@{\hspace{10mm}}c}{\rule[-2mm]{0mm}{5mm}Target} &
\multicolumn{1}{c}{$P_x^\prime/P_z^\prime$} &
Ref. \\
\hline
$^4$He & -0.862 $\pm$ 0.020 $\pm$ 0.03 & this work \\
$^1$H  & -0.978 $\pm$ 0.044 $\pm$ 0.03 & this work \\
$^1$H  & -0.952 $\pm$ 0.008 & \cite{Mil98,Jones,world-data} 
\end{tabular}
\label{table-ratio}
\end{table}
}

Table~\ref{table-ratio} lists the extracted polarization ratios for
$^4$He and $^1$H.  No radiative corrections have been applied; 
they are expected to have an effect which is of order 1\% \cite{Afanasev}.
The hydrogen ratio is found to be in agreement with the polarization ratio
derived from the world average
 of $G_E/G_M$ \cite{Mil98,Jones,world-data}
for data between $Q^{2}$ of 0.3 and 0.5 (GeV/c)$^{2}$.

Since systematic effects on the polarization ratio for hydrogen and
helium were nearly the same in both size and sign, the effect of systematic
uncertainties on the ratio of helium to hydrogen data

\begin{equation}
 R = (P_{x}^\prime/P_{z}^\prime)_{\rm He}\;/\;(P_{x}^\prime/P_{z}^\prime)_{\rm H}
\label{eq:superrat}
\end{equation}

\noindent nearly canceled.  This ``super-ratio'' is estimated to have
a systematic uncertainty of less than 0.01.  The uncertainty
on $R$ is then 
limited by the statistical uncertainty, mainly of the hydrogen ratio
measured in this experiment. Using the ratio of $G_E/G_M$ derived from the 
world average for hydrogen would give a smaller statistical uncertainty
but larger  systematic uncertainty. 
 Table~\ref{table-super-ratio} lists the super-ratio,
  using the hydrogen ratio from this experiment, as a
function of $p_m$, along with the value averaged over the
entire data set. Negative values
of $p_m$ 
correspond to the recoiling nucleus having a momentum component along the
direction of $\vec{q}$.

\vbox{
\begin{table}
\caption{$R$ as a function of missing momentum.  }
\begin{tabular}{@{\hspace{10mm}}r@{ }lll@{\hspace{10mm}}}
\multicolumn{2}{@{\hspace{10mm}}c}{\rule[-2mm]{0mm}{5mm}$p_m$ (MeV/c)} &
\multicolumn{1}{c}{$(P_{x}^\prime/P_{z}^\prime)_{\rm He}\;/\;(P_{x}^\prime/P_{z}^\prime)_{\rm H}$} & \\
\hline
\hspace{1mm}  -53 &  & 0.88 $\pm$ 0.05 $\pm$ 0.01 \\
\hspace{1mm}   55 &  & 0.89 $\pm$ 0.05 $\pm$ 0.01 \\
\hline
\multicolumn{2}{@{\hspace{.2in}}c}{mean} & 0.88 $\pm$ 0.04 $\pm$ 0.01 \\
\end{tabular}
\label{table-super-ratio}
\end{table}
}

A meaningful interpretation of the polarization ratio
measured for $^{4}$He with that of hydrogen can only be made by
utilizing theoretical
calculations which include the effects of FSI,
the off-shell current operator, 
relativistic effects, MEC, and IC  on the ratio.
In addition,  any calculation must be
averaged over the spectrometer acceptance.

We now proceed as follows.   The polarization
transfer is predicted using a model which includes 
the free form factors and the best phenomenologically determined
optical potentials and bound state wave functions (BSWF), and 
 FSI. MEC and IC are included in one nonrelativistic model.
  If the value predicted using the free form
factor  does not describe the
measured value well, within the theoretical uncertainties,
the effect of a modified form factor
will be considered.  If the new
value  predicted  provides a better description of the
data, we  can  take it   as evidence that the proton  form factors
inside $^{4}$He  differ from those of a free proton.

Figure~\ref{ratio1} shows a comparison of the experimental result
$R$, the $^4$He polarization ratio
normalized to the hydrogen ratio, with the
acceptance-averaged calculations. The hydrogen calculation made use of
the same form factor parameterization as does the corresponding $^4$He
calculation.
\vbox{
\begin{figure}[htb]
\centerline{\psfig{figure=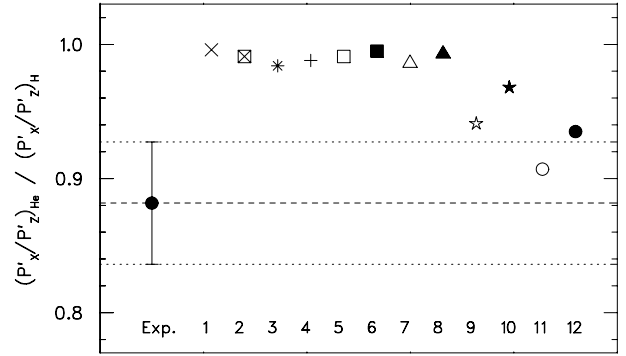,width=80mm,angle=90}}
\vskip0.4cm
\caption{Comparison of measured $R$ (Exp.) with
theoretical calculations. Laget - PWIA (1);  full calculation (2).
Udias - PWIA, $cc1$ (3),  $cc2$ (4); positive energy projection, 
$cc1$ (5), $cc2$(6);  
no spinor distortions, $cc1$ (7), $cc2$ (8);  
fully relativistic,  $cc1$ (9), $cc2$ (10);
fully relativistic, and QMC, $cc1$ (11), $cc2$ (12).}
\label{ratio1}
\end{figure}
}

We first examined the
effect of MEC and IC  using the non-relativistic calculations of 
Laget \cite{Laget94}.   The result of the
full Laget calculation was found to be nearly identical to the PWIA
result, points 2 and 1, respectively, in Fig.~\ref{ratio1}, 
indicating that MEC and IC do not contribute significantly
in our kinematics.  There is a  discrepancy 
of over two standard deviations  between the
observed value and both calculations of Laget.

We next use the model of Udias {\it et al.} to determine the magnitude
of relativistic effects.  Udias solves the Dirac equation and uses
relativistic optical potentials, 
but does not include MEC or IC.   For each case, we give the result
for two different de Forest \cite{def83} off-shell current operators,
$cc1$ and $cc2$.  The PWIA calculation of Udias (points 3 and 4)
includes positive and negative energy components for the bound state,
but only positive energy components for the ejected nucleon. It
gives nearly the same results as that of Laget, indicating that
$R$ is insensitive to the negative energy components of the bound state.
The calculations are insensitive to differences
between the two forms of the current operator.

The optical potential for $p + ^{3}$H was obtained by folding
a density-dependent empirical effective $pN$ interaction (EEI) with the
measured charge density for tritium.  Kelly and Wallace \cite{Kelly94} derived
an effective interaction for nucleon-nucleus scattering 
for  $9\leq A \leq 208$, designated IA2, 
in which spinor distortion is represented by density-dependent
modifications that are very similar to those of the EEI model fitted to
proton elastic and inelastic scattering data.  
Some parameters were adjusted to fit $p^{4}$He data yielding a better
fit to the proton elastic scattering data than any previous
optical potential.  Furthermore,
the dominant source of density dependence is consistent with the spinor
distortion employed by relativistic $(e,e'p)$ calculations.  We investigated
the sensitivity to final-state interactions by 
 using several other optical models and found variations of
$\pm 0.02$ in the polarization ratio.
 
In principle, the result should not depend on the gauge used.  However, the
calculations do show a small gauge dependence. We show the result
using the  Coulomb gauge, which gives nearly the same ratio as 
the Landau  gauge; the Weyl gauge
gives a larger ratio by  $0.04$.

The results of the Udias relativistic calculation projecting out
the negative energy sector (points 5 and 6) and with no spinor
distortion (points 7 and 8), called EMA-noSV in Ref.~\cite{Udias00}, 
are  also nearly the same as the 
PWIA calculation (points 3 and 4), demonstrating the small influence
of relativistic effects, other than the negative
energy components of the outgoing nucleon wave function,
which are not included in any of the calculations 1 to 8, and of FSI, on $R$.
The  fully relativistic calculations
 are shown as points 9 and 10.  The ratio decreases noticeably,
in particular with $cc1$,
but it remains slightly larger than the observed ratio.   Both results 
are between one and two standard deviations from the
observed value. 

\vbox{
\begin{figure}[htb]
\centerline{\psfig{figure=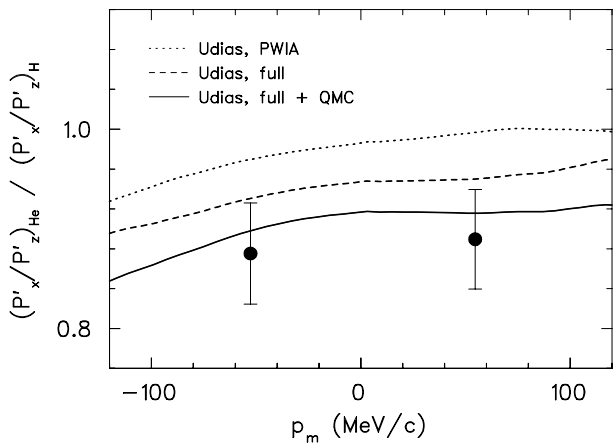,width=80mm,angle=90}}
\vskip0.4cm
\caption{$R$ as a function of missing momentum.  Using the
labels of Fig.~\ref{ratio1}, the curves are dotted  (3), 
dashed (9), and solid (11).}
\label{ratio2}
\end{figure}
}

Finally, we include the density dependent form factor modifications predicted 
by the QMC model of Lu {\it et al.} \cite{Adelaide}, using a bag
constant of 0.8 fm (points 11 and 12).
  These decrease the ratio further by about 4\%. 
The difference between the $cc1$ and $cc2$ results are about the same 
as for points 9 and 10.  The effect on the ratio is less than the 10\% effect
discussed in Ref.~\cite{Adelaide}. However, the
calculations  of Ref.~\cite{Adelaide} 
averaged over the bound state wavefunction.
As discussed in Ref.~\cite{Mal00}, an integration over the
final state, including the effects of absorption and non-locality
corrections, reduces the effect.  The addition of QMC brings the
calculation into  good agreement with the observed value.

Calculations using Kelly's EMA
for similar conditions give  very similar results.
 The variations in the result due to choice of the BSWF
 and the effect of Coulomb distortions were
 negligible in  both models.

The dependence of $R$ on the missing momentum was found to be small,
as shown in Fig.~\ref{ratio2}. A sample of the calculations
given in Fig.~\ref{ratio1} are shown in Fig.~\ref{ratio2}.  
The other calculations give curves 
nearly parallel to those shown, with separations about
the same as the separations of the averages shown in Fig.~\ref{ratio1}.  

In conclusion,
we have measured polarization 
transfer in the reaction $^4$He$(\vec{e},e'\vec{p})^3$H
for the first
time.  The $P_{x}^\prime/P_{z}^\prime$ ratio is in clear disagreement
with PWIA and  non-relativistic 
calculations.  A full relativistic calculation agrees
at the two standard deviation level.  The variation in
the result for different choices  of the bound-state
wave function, optical model, and current operator, added in
quadrature, is less
than one standard deviation. These measurements give the first 
evidence that a fully  
relativistic calculation that includes negative energy components
giving rise to spinor distortions,
is required for a correct description
of spin transfer in $(\vec{e},e'\vec{p})$ 
for $^{4}$He, even at low missing momentum.
The addition of a modified
proton form factor to the calculation, predicted by the QMC model, 
brings the result into good
agreement with the data.  Although the data do favor the models
with a modified form factor, the statistical significance is not
sufficient to exclude calculations without form factor modification.

We wish to thank the MAMI accelerator staff for their excellent
work during this experiment.
The authors gratefully acknowledge the assistance of  J.-M. Laget
 with theoretical calculations.  
This work was supported in part by the Deutsche Forschungsgemeinschaft,
U.S. National Science Foundation, and the Schweizerische Nationalfond. 

\noindent $^{a}$Permanent address: Department of Physics, University of Zagreb,
Zagreb, Croatia\\
$^{b}$Present address: Physik Department E18, T.U. M\"unchen, Germany\\
$^{c}$Present address: Netherlands Industrial Property Office, Rijswijk,
The Netherlands\\
$^{d}$Present address: Glasgow University, Glasgow, Scotland, UK\\

\end{document}